\documentclass[a4paper,12pt]{article}
\usepackage{jheppub}

\usepackage[T1]{fontenc}
\usepackage{CJK}
\usepackage{xcolor}
\usepackage{amsfonts}
\usepackage{epstopdf}
\usepackage{wrapfig} % to wrap figure with
\usepackage{subfigure}
\usepackage{graphicx}  % Include figure files
\usepackage{dcolumn}   % Align table columns
\usepackage{bm}
\usepackage{float}

\title{$SL(2,R)\times U(1)$ symmetry and quasinormal modes in the self-dual warped AdS black hole}
\author{Yuan Chen,}
\author{Wei Guo,}
\author{Kai Shi,}
\author{Hongbao Zhang}

\affiliation{Department of Physics, Beijing Normal University, Beijing 100875, China}
\emailAdd{yuanchen@mail.bnu.edu.cn}
\emailAdd{weiguo@mail.bnu.edu.cn}
\emailAdd{kaishi@mail.bnu.edu.cn}
\emailAdd{hongbaozhang@bnu.edu.cn}

\abstract{The algebraic approach to the spectrum of quasinormal modes
has been made as simple as possible for the BTZ black hole by the strategy developed in \cite{Zhang}. By working with the self-dual warped AdS black hole, we demonstrate in an explicit way that such a strategy can be well adapted to those warped AdS balck holes with the $SL(2,R)\times U(1)$ isometry. To this end,  we first introduce two associated tensor fields with the quadratic Casimir of $SL(2,R)\times U(1)$ Lie algebra in the self-dual warped AdS black hole and show that they correspond essentially to the metric and volume element up to a constant prefactor, respectively. Then without appealing to any concrete coordinate system, we can further show that the solutions to the equations of motion for the scalar, vector, spinor fields all fall into the representations of the $SL(2,R)\times U(1)$ Lie algebra by a purely abstract tensor and spinor analysis. Accordingly, the corresponding spectrum of quasinormal modes for each fixed azimuthal quantum number can be derived algebraically as the infinite tower of descendants of the highest weight mode of the $SL(2,R)$ Lie subalgebra.
}

\begin{document}
\maketitle
\flushbottom
\section{Introduction}
As we know, not only does the photon ring outside of the black hole play a vital role in predicting the intricate patterns of the black hole image, but also controls the photon sphere quasinormal ringdowns of massless fields such as the gravitational waves. Very recently, it has been discovered in \cite{Raffaelli} that the photon sphere quasinormal modes exhibit an emergent $SL(2,R)$ symmetry for the static spherically symmetry black hole, namely the spectrum of the photon sphere quasinormal modes forms the highest weight representation of the emergent $SL(2,R)$ symmetry, which is further found to persist also for the stationary rotating Kerr black hole\cite{HKLS}. Later on, to gain a deeper understanding of this emergent $SL(2,R)$ symmetry, the authors in \cite{KLS} initiate the exploration of the self-dual warped AdS black hole, which arises as an approximation to the near-extremal Kerr black hole with the $SL(2,R)$ isometry.
In particular, the resultant spectrum of quasinormal modes by explicit calculation can be identified as the highest weight representation of the $SL(2,R)$ isometry. It is further shown in \cite{CHH} that the aforementioned emergent $SL(2,R)$ symmetry associated with the spectrum of the photon sphere quasinormal modes, if redefined appropriately by virtue of the residual degrees of freedom, can be regarded precisely as the eikonal limit of the $SL(2,R)$ isometry.

As in the context of AdS/CFT correspondence, the identification of such a symmetry is suspected to offer us a guiding principle to search for the holographic dual field theory description of the black hole under consideration through Kerr/CFT correspondence and warped AdS/CFT correspondence, respectively\cite{GHSS,Compere,CZZZ,SX}. In particular, the $SL(2,R)$ symmetry dictated quasinormal modes correspond to the poles of the retarded Green function of the putative dual CFT. Such a specific correspondence has been confirmed in AdS$_3$/CFT$_2$, where analytic results can be readily obtained for both sides. Speaking specifically, the analytic result for the quasinormal modes of the scalar, vector, and spinor perturbations in the BTZ black hole is first obtained by directly solving the equations of motion\cite{CL,BSS}. Later on, inspired by \cite{MS} and \cite{BKL}, the authors in \cite{SS} demonstrate that the quasinormal modes can be constructed as the left and right
chiral highest weight representation of the $SL(2,R)$ isometry for the scalar and metric perturbations. This algebraic construction
has been further generalized to other three dimensional black holes with the
vector perturbation included\cite{CLC,CZ}. However, the involved
calculation is somewhat complicated, which makes
the $SL(2,R)$ isometry obscure albeit recovered in
the final result as it should be the case. Such a technical deficiency is rescued in \cite{Zhang} by resorting to the two intrinsic tensor fields associated with the $SL(2,R)$ isometry of the BTZ black hole as well as the covariant derivative rather than the ordinary derivative in the course of analysis. Among others, one very advantage of this strategy is that the spinor perturbation can be incorporated and treated in a uniform manner as other tensor fields.

As alluded to in the very beginning, the $SL(2,R)$ isometry of the self-dual warped AdS black hole is supposed to play a similar role in controlling the spectrum of quasinormal modes algebraically.
%which may help develop the link of the holographic perspective on gravity to observational gravitational-wave signatures.
%This may lead to some intriguing observational signatures of holographic gravity, since the quasinormal ringdowns,  potentially observable in the gravitational waveform,  admit a holographic interpretation as quantum Ruelle resonances of the thermal quantum state dual to the near extremal Kerr black hole\cite{HKLS,KLS}. 
However, such a suspicion has so far been confirmed solely for the (massless) scalar perturbation due to its obvious simplicity. So it is tempting to check this suspicion for other perturbations explicitly by utilizing the strategy developed in \cite{Zhang}. The purpose of this paper is to show this is also the case for both vector and spinor perturbations. To make the whole analysis as simple as possible, we are required to introduce the two intrinsic tensor fields associated with the Casimir of the full $SL(2,R)\times U(1)$ isometry of the self-dual warped AdS black hole. This is the main difference from the BTZ black hole, where the full isometry is given by $SL(2,R)_L\times SL(2,R)_R$ but the two intrinsic tensor fields are associated with the Casimir of each $SL(2,R)$ sector. It is noteworthy that the warp factor makes the additional $U(1)$ isometry also display itself in Riemann curvature, which turns out to have an extra effect on the conformal weight of the spinor perturbation. Except for these nuances, the whole procedure devised in \cite{Zhang} proves to be applicable to the self-dual warped AdS black hole with the $SL(2,R)\times U(1)$ symmetry manifest throughout the whole analysis.

The rest of paper is organized as follows. In the subsequent section, we provide a brief review of the self-dual warped AdS black hole as a solution to the topological massive gravity, where the $SL(2,R)\times U(1)$ isometry is identified and relevant geometric quantities are presented in an explicit way for our later usage. In Section \ref{Casimir}, with the quadratic Casimir of $SL(2,R)\times U(1)$ Lie algebra, we introduce its Lie derivative representation and two associated tensor fields, which turn out to be proportional to the metric and the volume element, respectively.
 In Section \ref{derivation}, we first derive how the solutions
to the equations of motion fall in a uniform manner into the representations of the $SL(2,R)\times U(1)$
Lie algebra for the scalar, vector, and spinor field in the self-dual warped AdS black hole and then present the algebraic approach to the quasinormal modes as the highest weight representation of the $SL(2,R)$ Lie subalgebra for each fixed azimuthal quantum number, where the main bulk of tensor and spinor analysis is performed in a purely abstract way without resorting to any specific coordinate system\footnote{To our best knowledge, the equation of motion for the metric perturbation on top of the warped AdS is essentially third order in the topological massive gravity, and has not been tamed into so well understood a form as in AdS. So we leave it to future work. }. We conclude our paper in the last section.

Notation and conventions follow \cite{Wald} unless specified
otherwise.
\section{Self-dual warped AdS black hole, $SL(2,R)\times U(1)$ isometry and relevant geometric quantities}
Let us start with the self-dual warped AdS black hole
\begin{equation}
ds^2=\ell^2\left[\frac{-dT^2+dx^2}{\sinh^2 x}+\Lambda^2\left(d\phi+\frac{dT}{\tanh x}\right)^2\right],
\end{equation}
where $\ell$ is the warped AdS radius and the warp factor $\Lambda$ is so defined that the unwarped self-dual AdS black hole is given by $\Lambda=1$.
%In what follows, we would like to work in the light cone
%coordinates, i.e., $u=\tau+\varphi, v=\tau-\varphi$, in which the
%metric takes the form
%\begin{equation}
%\begin{aligned}
%g_{ab}=&\ell^2\left[(\Lambda^2\coth^2 x- \text{csch}^2 x) (dT)_a (dT)_b+\Lambda^2\coth x[(dT)_a (d\phi)_b+(d\phi)_a (dT)_b] \right.\\&\left.+\text{csch}^2x (dx)_a (dx)_b+\Lambda^2(d\phi)_a(d\phi)_b\right]
%\end{aligned}
%\end{equation}
Whence the inverse metric can be obtained as
\begin{equation}
\begin{aligned}
g^{ab}=&\frac{1}{\ell^2}\left\{{-\sinh^2 x}(\frac{\partial}{\partial
T})^a(\frac{\partial}{\partial
T})^b+\cosh x \sinh x[(\frac{\partial}{\partial
T})^a(\frac{\partial}{\partial \phi})^b+(\frac{\partial}{\partial
\phi})^b(\frac{\partial}{\partial
T})^a]\right.\\
&\left.+\sinh^2 x(\frac{\partial}{\partial x})^a(\frac{\partial}{\partial
x})^a+\frac{1-\Lambda^2\cosh^2{x}}{\Lambda^2}(\frac{\partial}{\partial\phi})^a(\frac{\partial}{\partial\phi})^b\right\},
\end{aligned}
\end{equation}
and the associated volume element reads
\begin{equation}
\epsilon=\Lambda\ell^3\ \text{csch}^2 x\ dT\wedge dx\wedge d\phi.
\end{equation}

Such a black hole has the $SL(2,R)\times U(1)$ isometry with the corresponding Killing fields defined as follows
\begin{eqnarray}
L_0^a&=&-(\frac{\partial}{\partial T})^a, \nonumber\\
L_{-1}^a&=&e^{-T}[-\cosh{x}(\frac{\partial}{\partial
T})^a+\sinh{x}(\frac{\partial}{\partial
x})^a+\sinh{x}(\frac{\partial}{\partial\phi})^a],\nonumber\\
L_{+1}^a&=&e^{T}[-\cosh{x}(\frac{\partial}{\partial
T})^a-\sinh{x}(\frac{\partial}{\partial
x})^a+\sinh{x}(\frac{\partial}{\partial\phi})^a],\nonumber\\
W_0^a&=&(\frac{\partial}{\partial \phi})^a,\label{Killing}
\end{eqnarray}
whereby the commutators satisfy the following $SL(2,R)\times U(1)$ Lie algebra
\begin{equation}
[L_0,L_{\pm1}]=\mp L_{\pm1},\quad [L_{+1},L_{-1}]=2L_0,\quad [W_0,L_m]=0.
\end{equation}
with $m=0,\pm1$.
Note that our self-dual warped AdS black hole is not locally maximally symmetric. Instead, one can show that the corresponding Riemann tensor, Ricci tensors, Ricci scalar, and Einstein tensor are given by
\begin{equation}
\begin{aligned}
R_{abcd}&=2(g_{a[c}R_{d]b}-g_{b[c}R_{d]a})-Rg_{a[c}g_{d]b},\\
R_{ab}&=\frac{\Lambda^2-2}{2\ell^2}g_{ab}+\frac{1-\Lambda^2}{\Lambda^2\ell^4}W_{0a}W_{0b},\\
R&=\frac{\Lambda^2-4}{2\ell^2},\\
G_{ab}&=\frac{\Lambda^2}{4\ell^2}g_{ab}+\frac{1-\Lambda^2}{\Lambda^2\ell^4}{W_0}_a{W_0}_b.
\end{aligned}
\end{equation}
As one can see, the Riemann tensor receives an extra contribution from the $U(1)$ generator. In particular, this $U(1)$ generator will also leave its footprint in the following expression, i.e.,
\begin{equation}
R_{abcd}\epsilon^{ab}{}_e\epsilon^{cd}{}_f=\frac{\Lambda^2}{\ell^2}g_{ef}+\frac{4(1-\Lambda^2)}{\Lambda^2\ell^4}W_{0e}W_{0f},\label{first}
\end{equation}
which, as mentioned before and shown later, will have an effect on the conformal weight of the spinor perturbation. On the other hand, although the Weyl tensor vanishes automatically, the self-dual warped AdS black hole is not conformally flat because the Cotton tensor defined as $C_{ab}=\epsilon_a{}^{cd}\nabla_c(R_{db}-\frac{1}{4}Rg_{db})$ is given by
\begin{equation}
C_{ab}=\frac{\Lambda(1-\Lambda^2)}{2 \ell^3}g_{ab}+\frac{3(\Lambda^2-1)}{2\Lambda\ell^5}{W_0}_a{W_0}_b,
\end{equation}
which does not vanish when $\Lambda\neq1$. Accordingly, we have
\begin{equation}
    G_{ab}-\frac{4-\Lambda^2}{12\ell^2}g_{ab}+\frac{2\ell}{3\Lambda}C_{ab}=0,
\end{equation}
which means that our self-dual warped AdS black hole can be supported by the gravitational Chern-Simons term as a solution to the topological massive gravity\cite{KLS}.

\section{Quadratic Casimir, its Lie derivative representation and two associated tensor fields\label{Casimir}}
%Thus such a black hole also admits six Killing fields. A Killing
%field $\xi$, by definition, is a vector field which can generate
%one-parameter group of isometries, or equivalently, a vector field
%satisfying the Killing equation $\nabla_a\xi_b=\nabla_{[a}\xi_{b]}$
%with $\nabla_a$ the covariant derivative operator. With this, one
%can easily show that the Lie derivative with respect to any Killing
%field Kills all intrinsic tensor fields associated with the metric
%such as the volume element and commutes with the covariant
%derivative operator. Here we denote these six Killing fields by
%$L_k$ and $\bar{L}_k$ with $k=0,\pm1$. In particular, $L_k$ is given
%by

%Similarly, $\bar{L}_k$ is defined as (\ref{Killing}) simply by
%switching $u$ and $v$ therein. Locally their Lie commutators satisfy
%two sets of the $SL(2,R)$ Lie algebra, i.e.,
The quadratic Casimir operator of $SL(2,R)\times U(1)$ Lie algebra is defined as
\begin{equation}
    C^2=L^2+(1-\frac{1}{\Lambda^2})W_0^2
\end{equation}
with
\begin{equation}
L^2=L_0^2-\frac{1}{2}(L_{+1}L_{-1}+L_{-1}L_{+1})
\end{equation}
the Casimir operator of $SL(2,R)$ Lie subalgebra.
Note that the Lie derivative of tensor and spinor fields obeys
$[\mathcal{L}_X,\mathcal{L}_Y]=\mathcal{L}_{[X,Y]}$ and
$\mathcal{L}_{\alpha X}=\alpha\mathcal{L}_{X}$ for the arbitrary Killing
vector fields $X$ and $Y$ with the arbitrary constant
$\alpha$. Therefore
the above Lie algebra can be naturally represented by the Lie
derivative. In particular, the quadratic Casimir operators can be
realized by the Lie derivative as follows
\begin{equation}
\mathcal{C}^2=\mathcal{L}^2-(1-\frac{1}{\Lambda^2})\mathcal{L}_{W_0}\mathcal{L}_{W_0}
=\mathcal{L}_{L_0}\mathcal{L}_{L_0}-\frac{1}{2}(\mathcal{L}_{L_{+1}}\mathcal{L}_{L_{-1}}+\mathcal{L}_{L_{-1}}\mathcal{L}_{L_{+1}})-(1-\frac{1}{\Lambda^2})\mathcal{L}_{W_0}\mathcal{L}_{W_0}.\label{commutation}
\end{equation}

Now inspired by the strategy developed in \cite{Zhang}, we construct the following two tensor fields associated with the quadratic Casimir of $SL(2,R)\times U(1)$, i.e.,
\begin{equation}
H^{ab}=L_0^aL_0^b-\frac{1}{2}(L_{+1}^aL_{-1}^b+L_{-1}^aL_{+1}^b)-(1-\frac{1}{\Lambda^2})W_0^aW_0^b
\end{equation}
and
\begin{eqnarray}
Z_{abc}&=&L_{0a}\nabla_bL_{0c}-\frac{1}{2}(L_{+1a}\nabla_bL_{-1c}+L_{-1a}\nabla_bL_{+1c})-(1-\frac{1}{\Lambda^2}){W_0}_a\nabla_b{W_0}_c\nonumber.
\end{eqnarray}
Obviously, $H$ is a symmetric tensor field. By a straightforward calculation, one can show that $H$ is actually proportional to our metric, i.e.,
\begin{equation}
H^{ab}=\ell^2g^{ab}.\label{fH}
\end{equation}
Regarding $Z$, we first notice that it is
antisymmetric between the last two indices due to the
Killing equation $\nabla_a\xi_b=\nabla_{[a}\xi_{b]}$. Second,
$\nabla_bH_{ac}=0$ tells us that it is also
antisymmetric between the first and third indices. So
$Z$ is virtually a $3$-form, which implies that it should be proportional to the  volume element.
In addition, we have
\begin{eqnarray}
\nabla^aZ_{abc}&=&L_{0a}\nabla^a\nabla_bL_{0c}-\frac{1}{2}(L_{+1a}\nabla^a\nabla_bL_{-1c}+L_{-1a}\nabla^a\nabla_bL_{+1c})-(1-\frac{1}{\Lambda^2}){W_0}_a\nabla^a\nabla_b{W_0}_c \nonumber\\
&=&{R_{cb}}^{ad}[L_{0a}L_{0d}-\frac{1}{2}(L_{+1a}L_{-1d}+L_{-1a}L_{+1d})-(1-\frac{1}{\Lambda^2}){W_0}_a{W_0}_c]=0,
\end{eqnarray}
where $R_{abcd}=R_{ab[cd]}$ is used in the last step while
the Killing equation and the
identity
\begin{equation}
\nabla_a\nabla_b\xi_c={R_{cba}}^d\xi_d \label{second}
\end{equation}
for any Killing field $\xi$ are used in the first and second steps, respectively. Therefore, the proportional coefficients in
front of the volume element should be constant. In particular, an
explicit computation leads to
\begin{equation}
Z_{abc}=-\frac{\Lambda\ell}{2}\epsilon_{abc}.\label{fZ}
\end{equation}
It is noteworthy that the nice properties exhibited in Eq. (\ref{fH}) and Eq. (\ref{fZ}) respectively for $H$ and $Z$ fields will be lost if one naively constructs them simply out of the quadratic Casimir of $SL(2,R)$ Lie subalgebra.

\section{$SL(2,R)\times U(1)$ symmetry and quasinormal modes in the self-dual warped AdS black hole\label{derivation}}
As a warm-up, let us rework with the scalar field $\phi$, whose
equation of motion is given by
\begin{equation}
(\nabla_a\nabla^a-\mu^2)\phi=0.
\end{equation}
The Lie derivative acting on the scalar field gives rise to
\begin{equation}
\mathcal{L}_X\mathcal{L}_Y\phi=X^a\nabla_a(Y^b\nabla_b\phi)=(X^a\nabla_aY^b)\nabla_b\phi+X^aY^b\nabla_a\nabla_b\phi,
\end{equation}
whereby we have
\begin{equation}
\mathcal{C}^2\phi={{Z^a}_a}^b\nabla_b\phi+H^{ab}\nabla_a\nabla_b\phi=\ell^2g^{ab}\nabla_a\nabla_b\phi=(\mu\ell)^2\phi.
\end{equation}

Now let us proceed with the massive vector field $A$, whose equation of
motion is given by
\begin{equation}
{\epsilon_a}^{bc}\nabla_bA_c=-\mu A_a.
\end{equation}
Whence we can see that the Lorenz condition is satisfied automatically as follows
\begin{equation}
\nabla_aA^a=-\frac{1}{\mu}\epsilon^{abc}\nabla_a\nabla_bA_c=-\frac{1}{2\mu}\epsilon^{abc}{R_{abc}}^dA_d=-\frac{1}{2\mu}\epsilon^{abc}{R_{[abc]}}^dA_d=0
\end{equation}
due to the cyclic identity $R_{[abc]d}=0$. Furthermore, we have
\begin{eqnarray}
\mu^2A^d&=&-\mu\epsilon^{dea}\nabla_eA_a=\epsilon^{dea}\nabla_e({\epsilon_a}^{bc}\nabla_bA_c)=\epsilon^{ade}{\epsilon_a}^{bc}\nabla_e\nabla_bA_c
\nonumber\\
&=&(g^{dc}g^{eb}-g^{db}g^{ec})\nabla_e\nabla_bA_c=\nabla_a\nabla^aA^d-\nabla_a\nabla^dA^a
\nonumber\\
&=&\nabla_a\nabla^aA^d+\nabla^d\nabla_aA^a-\nabla_a\nabla^dA^a=\nabla_a\nabla^aA^d+R^{dabc}A_cg_{ab}
\nonumber\\
&=&\nabla_a\nabla^aA^d-R^{dc}A_c.
\end{eqnarray}
On the other hand, by the Lie derivative acting on this vector field, we have
\begin{eqnarray}
\mathcal{L}_X\mathcal{L}_YA_a&=&X^b\nabla_b(\mathcal{L}_YA_a)+\mathcal{L}_YA_b\nabla_aX^b
\nonumber\\
&=&X^b\nabla_b(Y^c\nabla_cA_a+A_c\nabla_aY^c)+(Y^c\nabla_cA_b+A_c\nabla_bY^c)\nabla_aX^b
\nonumber\\
&=&(X^b\nabla_bY^c)\nabla_cA_a+X^bY^c\nabla_b\nabla_cA_a+(X^b\nabla_aY^c)\nabla_bA_c+A_cX^b\nabla_b\nabla_aY^c\nonumber\\
&&+(Y^c\nabla_aX^b)\nabla_cA_b+A_c\nabla_b(Y^c\nabla_aX^b)-A_cY^c\nabla_b\nabla_aX^b.
\end{eqnarray}
Whence we can further obtain
\begin{eqnarray}
\mathcal{C}^2A_a&=&{{Z^b}_b}^c\nabla_cA_a+H^{bc}\nabla_b\nabla_cA_a+2{{Z^c}_a}^b\nabla_cA_b+A_c\nabla_b{{Z^c}_a}^b+A^cR_{cabd}H^{bd}-A_cR_{ad}H^{dc}
\nonumber\\
&=&\ell^2g^{bc}\nabla_b\nabla_cA_a+\ell \Lambda{\epsilon_a}^{cb}\nabla_cA_b-\ell^2R_{ac}A^c
\nonumber\\
&=&[(\mu\ell)^2-\Lambda \mu\ell]A_a,
\end{eqnarray}
where the identity (\ref{second}) is used in the first step.

With the above experience, let us manipulate the spinor field, which is more involved. To this end, we start with the Dirac equation
\begin{equation}
(\gamma^a\nabla_a+\mu)\psi=0.
\end{equation}
Here $\gamma^a=e_I^a\gamma^I$ and the covariant derivative acting on the spinor field is given by
$\nabla_a=\partial_a+\frac{1}{4}\omega_{IJ a}\gamma^{IJ}$, where
$e_I^a$ constitute a set of orthogonal normal vector bases, and Gamma
matrices satisfy $\{\gamma^I,\gamma^J\}=2\eta^{IJ}$ with the spin
connection $\omega_{IJa}=e_{Ib}\nabla_ae_J^b$ and
$\gamma^{IJ}=\frac{1}{2}[\gamma^I,\gamma^J]$. Next acting on the Dirac equation with $\gamma^b\nabla_b-\mu$, we obtain
\begin{eqnarray}
0&=&(\gamma^b\nabla_b-\mu)(\gamma^a\nabla_a+\mu)\psi=(\gamma^a\gamma^b\nabla_a\nabla_b-\mu^2)\psi\nonumber\\
&=&(g^{ab}\nabla_a\nabla_b+\gamma^{ab}\nabla_a\nabla_b-\mu^2)\psi=(\nabla_a\nabla^a-\mu^2)\psi+\gamma^{ab}\nabla_{[a}\nabla_{b]}\psi\nonumber\\
&=&(\nabla_a\nabla^a-\mu^2+\frac{1}{8}R_{abcd}\gamma^{ab}\gamma^{cd})\psi,
\end{eqnarray}
where $\gamma^{ab}=e_I^ae_J^b\gamma^{IJ}$.
On the other hand, the Lie derivative of spinor fields with respect to Killing fields is given by\cite{CD}
\begin{equation}
\mathcal{L}_X\psi=X^a\nabla_a\psi-\frac{1}{4}\gamma^{ab}\psi\nabla_bX_a,
\end{equation}
whereby we have
\begin{eqnarray}
\mathcal{L}_X\mathcal{L}_Y\psi&=&X^a\nabla_a\mathcal{L}_Y\psi-\frac{1}{4}\gamma^{ab}\mathcal{L}_Y\psi\nabla_bX_a\nonumber\\
&=&X^a\nabla_a(Y^c\nabla_c\psi-\frac{1}{4}\gamma^{cd}\psi\nabla_dY_c)-\frac{1}{4}\gamma^{ab}(Y^c\nabla_c\psi-\frac{1}{4}\gamma^{cd}\psi\nabla_dY_c)\nabla_bX_a\nonumber\\
&=&(X^a\nabla_aY^c)\nabla_c\psi+X^aY^c\nabla_a\nabla_c\psi-\frac{1}{4}\gamma^{cd}\psi X^a\nabla_a\nabla_dY_c-\frac{1}{4}(X^a\nabla_dY_c)\gamma^{cd}\nabla_a\psi\nonumber\\
&&-\frac{1}{4}(Y^c\nabla_bX_a)\gamma^{ab}\nabla_c\psi+\frac{1}{16}\gamma^{ab}\gamma^{cd}\psi\nabla_d(Y_c\nabla_bX_a)-\frac{1}{16}\gamma^{ab}\gamma^{cd}\psi
Y_c\nabla_d\nabla_bX_a.\nonumber\\
\end{eqnarray}
Then it is not hard to show
\begin{eqnarray}
\mathcal{C}^2\psi&=&{{Z^a}_a}^c\nabla_c\psi+H^{ac}\nabla_a\nabla_c\psi-\frac{1}{4}\gamma^{cd}\psi
R_{cdae}H^{ae}-\frac{1}{2}{Z^a}_{dc}\gamma^{cd}\nabla_a\psi\nonumber\\
&&+\frac{1}{16}\gamma^{ab}\gamma^{cd}\psi\nabla_dZ_{cba}-\frac{1}{16}\gamma^{ab}\gamma^{cd}\psi R_{abde}{H^e}_c\nonumber\\
&=&\ell^2\nabla_a\nabla^a\psi+\frac{\Lambda\ell}{4}{\epsilon^a}_{bc}\gamma^{cb}\nabla_a\psi+\frac{\ell^2}{16}R_{abcd}\gamma^{ab}\gamma^{cd}\psi\nonumber\\
&=&\ell^2(\mu^2-\frac{1}{16}R_{abcd}\gamma^{ab}\gamma^{cd})\psi-\frac{\Lambda\ell}{4}{\epsilon^a}_{bc}\gamma^{bc}\nabla_a\psi\nonumber\\
&=&\ell^2(\mu^2-\frac{1}{16}R_{abcd}\epsilon^{abe}\epsilon^{cdf}\gamma_e\gamma_f)\psi+\frac{\Lambda\ell}{2}\gamma^a\nabla_a\psi\nonumber\\
&=&[(\mu\ell)^2-\frac{\Lambda\mu\ell}{2}]\psi-(\frac{\Lambda^2}{16}g_{ef}+\frac{1-\Lambda^2}{4\Lambda^2\ell^2}W_eW_f)\gamma^{(e}\gamma^{f)}\psi\nonumber\\
&=&[(\mu\ell)^2-\frac{\Lambda\mu\ell}{2}]\psi-(\frac{3\Lambda^2}{16}+\frac{1-\Lambda^2}{4})\psi\nonumber\\
&=&[(\mu\ell)^2-\frac{\Lambda\mu\ell}{2}+\frac{\Lambda^2-4}{16}]\psi,
\end{eqnarray}
where the identity $\gamma^{ab}=\epsilon^{abc}\gamma_c$ and Eq. (\ref{first}) have been used in the fourth and fifth steps, respectively.
%Similarly, we have
%\begin{equation}
%\bar{\mathcal{L}}^2\psi=\frac{1}{4}(m^2-\frac{3}{4}-m)\psi.
%\end{equation}

So the upshot of the whole bulk of tensor and spinor analysis presented above is that the solutions to the equations of motion for various fields all turn out to fall into the representations of the $SL(2,R)\times U(1)$ Lie algebra characterized by the value of the Casimir, i.e.,
\begin{equation}
    \mathcal{C}^2\Phi=\lambda\Phi
\end{equation}
with $\lambda=(u\ell)^2$ for the scalar field, $\lambda=(u\ell)^2-\Lambda\mu\ell$ for the vector field, and $\lambda=(u\ell)^2-\frac{\Lambda\mu\ell}{2}+\frac{\Lambda^2-4}{16}$ for the spinor field.
%\section{Quasinormal modes in the self-dual AdS black hole}
%As a recapitulation, we find that the solutions to the equations of
%motion for various fields fall into the various representations of
%$SL(2,R)$ Lie algebra labeled by the value of the Casimir, i.e.,
%\begin{equation}
%\mathcal{L}^2\Phi=\lambda_+\Phi,\bar{\mathcal{L}}^2\Phi=\lambda_-\Phi,
%\end{equation}
%where $\lambda_\pm=\frac{m^2}{4}$ for the scalar field,
%$\lambda_\pm=\frac{m^2\pm 2m}{4}$ for the vector field,
%$\lambda_\pm=\frac{m^2\pm 4m+3}{4}$ for the tensor field, and
%$\lambda_\pm=\frac{m^2\pm m-\frac{3}{4}}{4}$ for our spinor field.

With this in mind, we can construct the quasinormal modes by the standard algebraic approach. Namely, we start
from the highest weight mode of the $SL(2,R)$ Lie subalgebra with a fixed azimuthal quantum number $m$ as follows
\begin{equation}
\mathcal{L}_{W_0}\Phi^{(m0)}=im\Phi^{(m0)},\quad
\mathcal{L}^2\Phi^{(m0)}=\lambda_L\Phi^{(m0)},\quad
\mathcal{L}_{L_{+1}}\Phi^{(m0)}=0,\quad
\mathcal{L}_{L_0}\Phi^{(m0)}=h\Phi^{(m0)},\label{rightweight}
\end{equation}
where the component of $\Phi^{m0}$ associated with the orthogonal normal basis chosen in Appendix B can be written formally as
\begin{equation}
    \Phi^{(m0)}=e^{-hT+im\phi}\Psi^{(m0)}(x)
\end{equation}
with the conformal weight $h$ and the eigenvalue $\lambda_L$ of the $SL(2,R)$ Casmir given respectively by
\begin{equation}
h=\frac{1\pm\sqrt{1+4\lambda_L}}{2},\quad
 \lambda_L=\lambda-\frac{(\Lambda^2-1)m^2}{\Lambda^2}.\label{irrelevant}
\end{equation}
Then the quasinormal modes can be obtained as the infinite tower of the descendent modes, i.e.,
\begin{equation}
    \Phi^{(mn)}=\mathcal{L}_{L_{-1}}^n\Phi^{(m0)}
\end{equation}
with $n=0, 1, 2, \cdots$. Note that $\mathcal{L}_{L_0}\Phi^{(mn)}=(h+n)\Phi^{(mn)}$, so we have the following spectrum of quasinormal frequencies
\begin{equation}
    \omega_{mn}=-i(h+n)
\end{equation}
with the imaginary part of $\omega_{mn}$ required to be negative by definition.

We conclude this section by mentioning that in the eikonal regime $|m|\gg 1$, $\lambda$ in Eq. (\ref{irrelevant}) can be neglected. This amounts to saying that the effects from the mass and spin are both subleading in the eikonal limit, which substantiates the claim made in \cite{KLS}.
\section{Conclusion}

%Here is my naive argument. The definition for one geodesic to another in the phase space is supposed to be
%\begin{equation}
   % \mathcal{L}_{v_f}v_H^a=\alpha v_H^a
%\end{equation}
%where the generator vector field is given by $v_f^a=\Omega^{ab}\nabla_bf$ for any function $f$ on the phase space with $\Omega_{ab}$ the symplectic form, $H$ the Hamiltonian, and $\alpha$ a scalar. Note that such a generator vector field is always canonical in the sense that the Lie derivative $\mathcal{L}_{v_f}\Omega_{ab}=0$.
%With this in mind, one can show for $f=HT^2$ that
%\begin{equation}
  %  \mathcal{L}_{v_f}\Omega^{ab}\nabla_b H=\Omega^{ab}\mathcal{L}_{v_f}(dH)_b=\Omega^{ab}[d(\mathcal{L}_{v_f}H)]_b=\Omega^{ab}(d\{H,HT^2\})_b=2\Omega^{ab}\nabla_b(HT)=2v^a_{HT},
%\end{equation}
%which seems not to be proportional to $v^a_{H}$, whose integral curve gives rise to the geodesic in phase space.

%Maybe such an argument comes from my wrong definition.
Partially motivated by the emergent $SL(2,R)$ symmetry in the photon sphere quasinormal modes of the Kerr black hole, we have successfully obtained the analytic expression for the spectrum of quasinormal modes of the scalar, vector, and spinor fields in the exactly soluble self-dual warped AdS black hole in a uniform manner by fully exploiting its $SL(2,R)\times U(1)$ isometry. To achieve this, we have introduced the two tensor fields associated with the Casimir of the full $SL(2,R)\times U(1)$ Lie algebra and unveiled their pleasing relations to the metric and volume element respectively.
Then we show that the solutions to the equations of motion of the scalar, vector, and spinor fields all fall into the representations of the $SL(2,R)\times U(1)$ Lie algebra by our tensor and spinor analysis, where no specific coordinate system is used and the aforementioned two tensor fields make $SL(2,R)\times U(1)$ symmetry transparent in the whole analysis. The resultant spectrum of quasinormal modes can be further constructed as the highest weight representation of the $SL(2,R)$ Lie subalgebra.

Although the self-dual warped AdS black hole is more involved than the simplest BTZ black hole, our work demonstrates that the strategy previously developed in \cite{Zhang} for the algebraic approach to the spectrum of quasinormal modes in the BTZ black hole turns out to be utterly applicable to the self-dual warped AdS black hole. As we know, there are other three dimensional warped AdS black hole solutions with the $SL(2,R)\times U(1)$ isometry\cite{MCL,BC,ALPSS}
, where a variety of quasinormal modes have been calculated out mainly by solving the equations of motion analytically\cite{CX1,CX2,CN,CNX,LR}. Note that the key to making our algebraic approach work lies in the three properties of our self-dual AdS black hole exhibited in Eq. (\ref{first}), Eq. (\ref{fH}), and Eq. (\ref{fZ}), which are believed to hold also for other warped AdS black holes except that the prefactors in front of the metric, quadratic of the $U(1)$ generator and volume element may be varied. Thus with our present work, we are convinced that the derivation of these quasinormal modes can also be made as simple as possible by our algebraic approach.

\section*{Acknowledgements}
This work is partly supported by the National Key Research and Development Program of China with Grant No. 2021YFC2203001 and National Natural Science Foundation of China with Grant No. 12075026.

%The author is grateful to Bin Chen for his talk at KITPC, which
%sparks his dive into this project. In addition, he would like to
%take this opportunity to thank Ronggen Cai and Elias Kiritsis for
%their very help to make possible his attending the long term AdS/CFT
%programme at KITPC. This research was supported in part by the PKIP
%of Chinese Academy of Sciences with Grant No. KJCX2.YW.W10. It was
%also supported by a European Union grant
%FP7-REGPOT-2008-1-CreteHEPCosmo-228644.
\section*{Appendices}
\subsection*{A An explicit calculation of $Z$}
By $\nabla_a\xi_b=\frac{1}{2}(d\xi)_{ab}$ for any
Killing field $\xi$, we have
\begin{equation}
\begin{aligned}
&&Z_{abc}=Z_{[abc]}=\frac{1}{2}[L_{0[a}dL_{0bc]}-\frac{1}{2}(L_{+1[a}dL_{-1bc]}+L_{-1[a}dL_{+1bc]})-(1-\frac{1}{\Lambda^2}){W_0}_{[a}{dW_0}_{bc]}]\\
&&=\frac{1}{6}[L_{0}\wedge dL_{0abc}-\frac{1}{2}(L_{+1}\wedge
dL_{-1abc}+L_{-1}\wedge dL_{+1abc})-(1-\frac{1}{\Lambda^2}){W_0}\wedge {dW_0}_{abc}],
\end{aligned}
\end{equation}
where
\begin{eqnarray}
L_{0a}&=&g_{ab}L_{0}^b=-\ell^2[(\Lambda^2\coth^2x-\text{csch}^2x)(dT)_a+\Lambda^2\coth x (d\phi)_a ],\\
L_{-1a}&=&g_{ab}L_{-1}^b=-e^{-T}\ell^2\text{csch}\  x[(\Lambda^2-1)\coth x(dT)_a-(dx)_a+\Lambda^2(d\phi)_a],\nonumber\\
L_{+1a}&=&g_{ab}L_{-1}^b=-e^{T}\ell^2\text{csch}\  x[(\Lambda^2-1)\coth x(dT)_a+(dx)_a+\Lambda^2(d\phi)_a],\nonumber\\
\end{eqnarray}
and
\begin{eqnarray}
(dL_0)_{ab}&=&\ell^2\text{csch}^2x[\Lambda^2 (dx)_a\wedge(d\phi)_b-2(\Lambda^2-1)\coth{x} (dT)_a\wedge(dx)_b], \nonumber\\
(dL_{-1})_{ab}&=&e^{-T}\ell^2\text{csch}\ x[(2-\Lambda^2-\Lambda^2\cosh^2{x})\text{csch}^2x (dT)_a\wedge(dx)_b-\Lambda^2 (d\phi)_a\wedge(dT)_b\nonumber\\
&&+\Lambda^2\coth{x} (dx)_a\wedge(d\phi)_b], \nonumber\\
(dL_{+1})_{ab}&=&e^{T}\ell^2\text{csch}\ x[(2-\Lambda^2-\Lambda^2\cosh^2{x})\text{csch}^2\ x (dT)_a\wedge(dx)_b+\Lambda^2 (d\phi)_a\wedge(dT)_b\nonumber\\
&&+\Lambda^2\coth{x} (dx)_a\wedge(d\phi)_b], \nonumber\\
(dW_0)_{ab}&=&\ell^2\Lambda^2\text{csch}^2x (dT)_a\wedge(dx)_b.
\end{eqnarray}
With this, we can finally obtain
\begin{equation}
Z=-\frac{\Lambda^2\ell^4\text{csch}^2 x}{2}dT\wedge dx\wedge d\phi.
\end{equation}
%\subsection*{B A little bit of Clifford algebra}
%Firstly by the identity
%\begin{equation}
%[A,BC]=\{A,B\}C-B\{A,C\},
%\end{equation}
%we have
%\begin{equation}
%[\gamma^I,\gamma^M\gamma^N]=2(\eta^{IM}\gamma^N-\eta^{IN}\gamma^M),
%\end{equation}
%which further gives
%\begin{equation}
%[\gamma^I,\gamma^{MN}]=2(\eta^{IM}\gamma^N-\eta^{IN}\gamma^M).
%\end{equation}
%Next by the Jacobi identity, we have
%\begin{eqnarray}
%[\gamma^{IJ},\gamma^{MN}]&=&\frac{1}{2}[[\gamma^I,\gamma^J],\gamma^{MN}]=\frac{1}{2}([\gamma^I,[\gamma^J,\gamma^{MN}]]-[\gamma^J,[\gamma^I,\gamma^{MN}])\nonumber\\
%&=&2(\eta^{JM}\gamma^{IN}-\eta^{JN}\gamma^{IM}-\eta^{IM}\gamma^{JN}+\eta^{IN}\gamma^{JM}).
%\end{eqnarray}
\subsection*{B A little bit of spinor analysis}
Associated with the choice of the orthogonal normal bases as
$e_0^a=\frac{\sinh{x}}{\ell}(\frac{\partial}{\partial T})^a-\frac{\cosh{x}}{\ell}(\frac{\partial}{\partial \phi})^a$,$e_1^a=\frac{\sinh{x}}{\ell}(\frac{\partial}{\partial x})^a$,
and $e_2^a=\frac{1}{\Lambda\ell}(\frac{\partial}{\partial\phi})^a$, the non-vanishing
spin connections can be written as
\begin{equation}
\begin{aligned}
\omega_{01a}=-\omega_{10a}&=\frac{(2-\Lambda^2)\coth{x}}{2}(dT)_a-\frac{\Lambda^2}{2}(d\phi)_a,\\
\omega_{02a}=-\omega_{20a}&=-\frac{\Lambda\text{csch}\ x}{2}(dx)_a,\\
\omega_{12a}=-\omega_{21a}&=\frac{\Lambda\text{csch}\ x}{2}(dT)_a.
\end{aligned}
\end{equation}
Thus we have
\begin{equation}
\begin{aligned}
\mathcal{L}_{L_0}\Psi(x)&=L_0^a\nabla_a\Psi(x)-\frac{1}{4}\gamma^{ab}\Psi(x)\nabla_bL_{0a}\\
&=L_0^a\partial_a\Psi(x)+\frac{1}{4}L^a_0\omega_{aIJ}\gamma^{IJ}\Psi(x)-\frac{1}{4}\gamma^{IJ}\Psi(x)e_I^ae_J^b\nabla_{b}L_{0a}\\
&=\frac{1}{4}\gamma^{IJ}\Psi(x)(L_0^a\omega_{aIJ}-e_I^ae_J^b\nabla_{b}L_{0a})
=0.
\end{aligned}
\end{equation}
Similarly, we can also obtain $\mathcal{L}_{W_0}\Psi(x)=0$.

\end{document}